\RequirePackage{fix-cm}

\documentclass[12pt,eqsecnum,a4paper]{article}
\usepackage{latexsym}
\usepackage{amsmath}
\usepackage{amssymb}
\usepackage{ulem}
\usepackage{fancyhdr}
\usepackage{url}
\usepackage{grffile}
\usepackage[vcentering,dvips]{geometry}
\usepackage{sidecap}
\usepackage{graphicx}
\usepackage{placeins}
\usepackage{booktabs,xltabular}
\usepackage{float}

\newcommand{\be}{\begin{equation}}
\newcommand{\ee}{\end{equation}}
\newcommand{\beq}{\begin{eqnarray}}
\newcommand{\eeq}{\end{eqnarray}}

\newcommand{\ba}{\begin{align}}
\newcommand{\ea}{\end{align}}

\begin{document}

\title{Effect of a viscous 
fluid shell on the propagation of
	gravitational waves
}


\author{
        Nigel~T. Bishop   \and
        Petrus~J. van der Walt
        \and
        Monos Naidoo 
}

\date{}

\maketitle
	\text{\small Department of Mathematics, Rhodes University, Grahamstown, 6140, South Africa} 
\begin{abstract}
In this paper we show that there are circumstances in which the damping of gravitational waves (GWs) propagating through a viscous fluid can be highly significant; in particular, this applies to Core Collapse Supernovae (CCSNe). In previous work, we used linearized perturbations on a fixed background within the Bondi-Sachs formalism, to determine the effect of a dust shell on GW propagation. Here, we start with the (previously found) velocity field of the matter, and use it to determine the shear tensor of the fluid flow. Then, for a viscous fluid, the energy dissipated is calculated, leading to an equation for GW damping. It is found that the damping effect agrees with previous results when the wavelength $\lambda$ is much smaller than the radius $r_i$ of the matter shell; but if $\lambda\gg r_i$, then the damping effect is greatly increased.

Next, the paper discusses an astrophysical application, CCSNe. There are several different physical processes that generate GWs, and many models have been presented in the literature. The damping effect thus needs to be evaluated with each of the parameters $\lambda,r_i$ and the coefficient of shear viscosity $\eta$, having a range of values. It is found that in most cases there will be significant damping, and in some cases that it is almost complete. 

We also consider the effect of viscous damping on primordial gravitational waves (pGWs) generated during inflation in the early Universe. Two cases are investigated where the wavelength is either much shorter than the shell radii or much longer; we find that there are conditions that will produce significant damping, to the extent that the waves would not be detectable.		
	\end{abstract}
	
	\maketitle
	
	
	\section{INTRODUCTION}
	\label{intro}
	
	We have shown, previously, that a dust shell surrounding a GW event modifies a gravitational (GW) wave in both magnitude and phase~\cite{Bishop:2019eff}, and extended the analysis to show that a burst of GWs can, in principle, lead to echoes~\cite{Naidoo:2021}, although, in practice, an astrophysical scenario that would produce a discernible echo is unlikely. We further showed that the results are astrophysically relevant: the GW signal from events including core collapse supernovae (CCSNe) and binary neutron star (BNS) mergers can be changed so that the modification is measurable~\cite{Naidoo:2021}. A key point about these effects is that the magnitude is proportional to $\lambda/r_i$, where $\lambda$ is the GW wavelength and $r_i$ is the matter shell radius.
	
	GWs travelling through a perfect fluid do not experience any absorption or dissipation~\cite{Ehlers1987} as noted in~\cite{Lu:2018smr}. However, Hawking ~\cite{Hawking1966} showed that in the case of nonzero shear viscosity, $\eta$, GWs travelling through such a fluid would interact with the matter losing energy to the medium and energy dissipation would occur, with a damping rate proportional to $16\pi\eta$; see also~\cite{Madore1973,Prasanna1999,Goswami2017}.
	
	In this paper, we calculate the effect of a viscous matter shell on GW propagation. The procedure is straightforward, but the intermediate algebraic expressions involve many terms, and so the calculation is handled using computer algebra. The starting point is a solution to the Einstein equations linearized about Minkowski in Bondi-Sachs form. The velocity field of dust in this spacetime is found, and from there it is a straightforward calculation to find the fluid shear tensor $\sigma_{ab}$, and thence the rate of energy dissipation due to viscosity. The result obtained reduces to previous results when $r_i\gg\lambda$, see Eq.~(\ref{h1}). However, when $\lambda\gg r_i$ the damping effect can be large, see Eq.~(\ref{h2}); to our knowledge, this case has not been considered previously.
	
	We next apply Eq.~(\ref{h2}) to the astrophysical case of CCSNe.
	CCSNe have long been regarded as a potential GW source for detection by LIGO/Virgo, but to date no such events have been observed. The modeling of CCSNe involves many aspects of matter physics; we review this literature, and so determine a value range for each parameter used in our model. The wavelength range is 300 to 3000\,km, corresponding to 100 to 1000\,Hz, and the source region radius range is 10 to 30\,km so that Eq.~(\ref{h2}) can be applied. There are reported values of the shear viscosity that lead to almost complete damping of the GW signal, whereas if the viscosity is somewhat lower then only the lower frequencies in the GW signal would be damped out.

As a further case study, we consider primordial gravitational waves (pGWs) generated during the epoch of cosmic inflation. Here, we set up an early Universe scenario where a discrete source generates pGWs, which are subsequently affected by viscous matter under the extreme conditions of the early Universe. Two cases are considered where the wavelength is either much shorter than the shell radii or alternatively much longer. In both cases there are conditions that will produce significant damping, to the extent that the waves will not be detectable.	
	
	In Section~\ref{theory} we discuss perturbations about Minkowski spacetime in the Bondi-Sachs formalism, and find the velocity field and then the shear tensor of the fluid flow. We then consider energy loss due to shear viscosity in Section~\ref{eloss}. The consequences to astrophysics are considered, in general, in Section~\ref{consider}. We consider the effects of a viscous fluid shell on the propagation of GWs in CCSNe in Section~\ref{ccsne}, and in Section~\ref{s-pGWs} we investigate the effects on pGWs. We summarise our results and conclusions in Section~\ref{summary}.
	
	We use geometric units in this paper, with the gravitational constant $G$ and the speed of light $c$ set to unity. However, results from the astrophysical literature are reported in SI units, and then used with the appropriate conversion factors.
	
	\section{USING BONDI-SACHS FORMALISM}
	\label{theory}
	
	We use the formalism developed previously~\cite{Bishop:2019eff,Naidoo:2021} on GWs propagating through matter shells. The metric is in Bondi-Sachs form
	\begin{align}
		ds^2  = & -\left(e^{2\beta}\left(1 + rW_c\right)
		- r^2h_{AB}U^AU^B\right)du^2
		- 2e^{2\beta}dudr \nonumber \\
		& - 2r^2 h_{AB}U^Bdudx^A
		+  r^2h_{AB}dx^Adx^B\,,
		\label{eq:bmet}
	\end{align}
	where $h^{AB}h_{BC}=\delta^A_C$, and the condition that $r$ is a surface area coordinate implies $\det(h_{AB})=\det(q_{AB})$ where $q_{AB}$ is a unit sphere metric (e.g. $d\theta^2+\sin^2\theta d\phi^2$). We represent $q_{AB}$ by a complex dyad (e.g. $q^A=(1,i/\sin\theta)$) and introduce the complex differential angular operators $\eth,\bar{\eth}$~\cite{Newman-Penrose-1966}, with the operators defined with respect to the unit sphere as detailed in~\cite{Bishop2016a,Gomez97}. Then $h_{AB}$ is represented by the complex quantity $J=q^Aq^Bh_{AB}/2$ (with $J=0$ characterizing spherical symmetry), and we also introduce the complex quantity $U=U^Aq_A$.
	We make the ansatz of a small perturbation about Minkowski spacetime with the metric quantities $\beta,U,W,J$ taking the form
	\begin{align}
		\beta=&\Re(\beta^{[2,2]}(r)e^{i\nu u}){}_0Z_{2,2}\,,\;\;
		U=\Re(U^{[2,2]}(r)e^{i\nu u}){}_1Z_{2,2}\,,\;\;
		W_c=\Re(W_c^{[2,2]}(r)e^{i\nu u}){}_0Z_{2,2}\,,\nonumber \\
		J=&\Re(J^{[2,2]}(r)e^{i\nu u}){}_2Z_{2,2}\,,
		\label{e-ansatz}
	\end{align}
	The perturbations oscillate in time with frequency $\nu/(2\pi)$. The quantities ${}_s Z_{\ell,m}$ are spin-weighted spherical harmonic basis functions related to the usual ${}_s Y_{\ell,m}$ as specified in~\cite{Bishop-2005b,Bishop2016a}. They have the property that ${}_0 Z_{\ell,m}$ are real, enabling the description of the metric quantities $\beta,W$ (which are real) without mode-mixing; however, for $s\ne 0$ ${}_s Z_{2,2}$ is, in general, complex. A general solution may be constructed by summing over the $(\ell,m)$ modes, but that is not needed here, since we are considering a source that is continuously emitting GWs at constant frequency dominated by the $\ell=2$ (quadrupolar) components.
	
	As shown in previous work~\cite{Bishop-2005b,Bishop:2019eff}, solving the vacuum Einstein equations under the condition of no incoming radiation leads to
	\begin{align}
		\beta^{[2,2]}=&b_0\,, \nonumber \\
		W_c^{[2,2]}=&4i\nu b_0-2\nu^4 C_{40}-2\nu^2 C_{30}+\frac{4 i \nu C_{30}-2b_0+4i\nu^3C_{40}}{r}+\frac{12 \nu^2 C_{40}}{r^2}\nonumber \\
		-&\frac{12i\nu C_{40}}{r^3}-\frac{6C_{40}}{r^4}
		\,,\nonumber \\
		U^{[2,2]}=&\frac{\sqrt{6}(-2 i\nu b_0+\nu^4C_{40}+\nu^2 C_{30})}{3}+\frac{2\sqrt{6} b_0}{r}
		+\frac{2\sqrt{6} C_{30}}{r^2}-\frac{4i\nu\sqrt{6} C_{40}}{r^4}
		\nonumber \\
		-&\frac{3\sqrt{6} C_{40}}{r^4} 
		\,,\nonumber \\
		J^{[2,2]}=&\frac{2\sqrt{6}(2b_0+i\nu^3C_{40}+i\nu C_{30})}{3}+\frac{2\sqrt{6}C_{30}}{r}
		+\frac{2\sqrt{6}C_{40}}{r^3}\,,
		\label{e-pert}
	\end{align}
	with constants of integration $b_0, C_{30}, C_{40}$. The gravitational news ${\mathcal N}$ is defined in a coordinate system that satisfies the Bondi gauge conditions $\lim_{r\rightarrow\infty}J,U,\beta,W/r =0$, and is calculated on making the required coordinate transformation. The procedure in the general case was described in~\cite{Bishop97b}. This was then simplified for the linearized approximation in~\cite{Bishop-2005b} (Sec.3.3), with an explicit expression for the news given in~\cite{Reisswig:2006}, Eq.~(16). Denoting the news for the solution Eq.~(\ref{e-pert}) by ${\mathcal N}_0$, and allowing for the conventions used here, we find ${\mathcal N}_{0}=-\sqrt{6}\nu^3 \Re(iC_{40}\exp(i\nu u))\,{}_2Z_{2,2}$. The rescaled gravitational wave strain and news are related (see~Eq.~(276) in \cite{Bishop2016a}) $\mathcal{H}_{0}=r(h_+ +ih_\times) = 2\int{\mathcal N}_{0} du$, giving
	\begin{equation}
		{\mathcal H}_{0}=\Re(H_{0} \exp(i\nu u))\,{}_2Z_{2,2}\;\mbox{with}\;\; H_{0}=-2\sqrt{6}\nu^2 C_{40}\,.
		\label{e-HM0}
	\end{equation}
	Thus, $C_{40}$ is determined by the physical problem being modelled, and $b_0,C_{30}$ represent gauge freedoms.
	
	We now suppose that the GWs pass through a shell of matter, and determine the velocity field $V_a$ treating the matter as dust. The ansatz for $V_a$ is similar to that for the metric, and is
	\begin{align}
		V_0&=-1+\Re(V^{[2,2]}_0(r)e^{i\nu u}){}_0Z_{2,2}\,,\;
		V_1=-1+\Re(V^{[2,2]}_1(r)e^{i\nu u}){}_0Z_{2,2}\,,\nonumber \\
		q^A V_A&=\Re(V^{[2,2]}_{ang}(r)e^{i\nu u}){}_1Z_{2,2}\,.
	\end{align}
	Then solving the matter conservation condition $\nabla_a(\rho v^aV^b)=0$, which in this case is equivalent to the geodesic condition, leads to
	\begin{align}
		&V^{[2,2]}_0(r)=\frac{1}{r^3}\times\nonumber \\
		&\left(3 C_{40} - 2 ir^3  \nu C_{30} - 2 ir^3  \nu^3  C_{40} + 6i \nu C_{40} r - 2i \nu b_0 r^4  + \nu^2  r^4  C_{30} + \nu^4  r^4  C_{40} - 6 C_{40} \nu^2  r^2 \right)\nonumber \\
		&V^{[2,2]}_1(r)=i\frac{9 C_{40} + 12 i\nu C_{40} r + 2i \nu b_0 r^4  - \nu^2  r^4  C_{30} - \nu^4  r^4  C_{40} - 6 C_{40} \nu^2  r^2 }{r^4\nu}\nonumber\\
		&V^{[2,2]}_{ang}(r)=-\frac{ i\sqrt{6}}{r^3\nu}\times\nonumber\\
		&\left(3 C_{40} - 2 ir^3  \nu C_{30} - 2 ir^3  \nu^3  C_{40} + 6i \nu C_{40} r - 2i \nu b_0 r^4  + \nu^2  r^4  C_{30} + \nu^4  r^4  C_{40} - 6 C_{40} \nu^2  r^2\right)\,.
		\label{e-V}
	\end{align}
	
	The contribution of viscosity to the stress-energy tensor of a viscous fluid is (see~\cite{Baumgarte2010a}, p.139)
	\begin{equation}
		T_{ab}=-2\eta\sigma_{ab}-\zeta\theta P_{ab}\,,
	\end{equation}
	where $\eta,\zeta$ are the coefficients of shear and bulk viscosity; $\theta$ is the fluid expansion, $\sigma_{ab}$ is the shear tensor, and $P_{ab}$ is the projection tensor, given by
	\begin{align}
		\theta&=g^{ab}\nabla_a V_b\,,\;\; P_{ab}=g_{ab}+V_a V_b\,,\nonumber \\
		\sigma_{ab}&=\frac{(P_{ac}\nabla_d V_b +P_{bc}\nabla_dV_a)g^{cd}}{2}-\frac{P_{ab}\theta}{3}\,.
	\end{align}
	The velocity field of Eq.~(\ref{e-V}) gives
	\begin{align}
		\theta&=\sigma_{00}=\sigma_{01}=\sigma_{0A}=0\,,\nonumber \\
		-\sigma^{[2,2]}_{11}&=\sigma^{[2,2]}_{W}=
		12 C_{40}\frac{3i-3 r \nu-ir^2\nu^2}{r^5\nu}\nonumber \\
		\sigma^{[2,2]}_{1U}&=
		2C_{40}\frac{6i-6r\nu-3ir^2\nu^2+r^3\nu^3}{r^4\nu}\nonumber \\
		\sigma^{[2,2]}_{J}&=
		C_{40}\frac{-3-3ir\nu +3r^2\nu^2+2ir^3\nu^3-r^4\nu^4}{r^3\nu}\,,
		\label{e-sig}
	\end{align}
	where the the above quantities are defined in terms of the usual separation of variables, i.e.,
	\begin{align}
		\sigma_{11}&=\Re(\sigma^{[2,2]}_{11}(r)e^{i\nu u}){}_0Z_{2,2}\,,
		q^A\sigma_{1A}=\Re(\sigma^{[2,2]}_{1U}(r)e^{i\nu u}){}_1Z_{2,2}\,,\nonumber \\
		q^{AB}\sigma_{AB}&=\Re(\sigma^{[2,2]}_{W}(r)e^{i\nu u}){}_0Z_{2,2}\,,
		q^Aq^B\sigma_{AB}=\Re(\sigma^{[2,2]}_{J}(r)e^{i\nu u}){}_2Z_{2,2}\,.
	\end{align}
	It is interesting to note that the expressions in Eqs.~(\ref{e-sig}) involve only the physical constant $C_{40}$, and not the gauge freedom constants $b_0,C_{30}$. Thus $\sigma_{ab}$ is gauge independent.
	
	\section{ENERGY LOSS DUE TO VISCOSITY}
	\label{eloss}
	
	We use the formula that the rate of energy loss per unit volume is $-2\eta \sigma_{ab}\sigma^{ab}$ where $\eta$ is the coefficient of shear viscosity~\cite{Baumgarte2010a}. This quantity is evaluated using Eqs.~(\ref{e-sig}), and then integrated over a shell of radius $r$ and thickness $\delta r$; the integration is straightforward because of the orthonormality of the angular basis functions ${}_sZ_{\ell,m}$. We find
	\begin{equation}
		\left<\dot{E}_\eta\right> = -12\eta C_{40}^2\nu^6 \delta r
		\left(1+\frac{2}{r^2\nu^2}+\frac{9}{r^4\nu^4}
		+\frac{45}{r^6\nu^6}+\frac{315}{r^8\nu^8}
		\right)\,,
		\label{e-dE_eta}
	\end{equation}
	where $\left<f\right>$ denotes the average of $f(u)$ over a wave period, i.e.
	\begin{equation}
		\left<f\right>=\frac{\nu}{2\pi}\int_0^{\frac{2\pi}{\nu}}fdt\,,
	\end{equation}
	and where we have used $\left<\cos^2(\nu u)\right>=\left<\sin^2(\nu u)\right>=1/2$ and $\left<\cos(\nu u)\sin(\nu u)\right>=0$.
	Now, the rate of energy being output in GWs is
	\begin{equation}
		\left<\dot{E}_{GW}\right>=\frac{1}{4\pi}\oint\left|N\right|^2
		=\frac{3C_{40}^2\nu^6}{4\pi}\,,
	\end{equation}
	so that
	\begin{equation}
		\left<\dot{E}_\eta\right> = -16\pi\eta \delta r \left<\dot{E}_{GW}\right>
		\left(1+\frac{2}{r^2\nu^2}+\frac{9}{r^4\nu^4}
		+\frac{45}{r^6\nu^6}+\frac{315}{r^8\nu^8}
		\right)\,.
	\end{equation}
	Conservation of energy implies that the energy being absorbed by the viscous fluid must be balanced by a reduction in the GW energy. Thus
	\begin{equation}
		\left<\dot{E}_{GW}\right>(r+\delta r)=\left<\dot{E}_{GW}\right>(r)
		\left[1-16\pi\eta \delta r 
		\left(1+\frac{2}{r^2\nu^2}+\frac{9}{r^4\nu^4}
		+\frac{45}{r^6\nu^6}+\frac{315}{r^8\nu^8}
		\right)\right]\,.
	\end{equation}
	As introduced earlier, $H$ represents the magnitude of the GWs rescaled to allow for the $1/r$ fall-off, i.e. $H\propto r \left<|h_+,h_\times|\right>$. Then $ \left<\dot{E}_{GW}\right> \propto H^2$ so that
	\begin{equation}
		H(r+\delta r)=H(r) \left[1-8\pi\eta \delta r 
		\left(1+\frac{2}{r^2\nu^2}+\frac{9}{r^4\nu^4}
		+\frac{45}{r^6\nu^6}+\frac{315}{r^8\nu^8}
		\right)\right]\,.
	\end{equation}
	This leads to the differential equation
	\begin{equation}
		\frac{dH}{dr}=-8\pi\eta H 
		\left(1+\frac{2}{r^2\nu^2}+\frac{9}{r^4\nu^4}
		+\frac{45}{r^6\nu^6}+\frac{315}{r^8\nu^8}
		\right)\,,
	\end{equation}
	which is easily solved to give
	\begin{equation}
		H(r)= C\exp\left(-8\pi\eta\left(r-\frac{2}{r\nu^2}-\frac{3}{r^3\nu^4}
		-\frac{9}{r^5\nu^6}-\frac{45}{r^7\nu^8}\right)\right)\,,
		\label{e-h0}
	\end{equation}
	where $C$ is a constant. There are two useful special cases. Let $r_i,r_o$ be the inner and outer radii of the shell. If $r_i,r_o$ are much larger than the wavelength $\lambda$ of the GWs, then
	\begin{equation}
		H(r_o)=H(r_i)\exp\left(-8\pi\eta (r_o-r_i)\right)\,.
		\label{h1}
	\end{equation}
	Results equivalent to Eq.~(\ref{h1}) have been given before, and some of the literature is reviewed in the Introduction, Section~\ref{intro}.
	If $r_i$ is much smaller than the wavelength of the GWs with $r_o=\alpha r_i$ with $\alpha>1$ then
		\begin{equation}
		H(r_o)=H(r_i)\exp\left(-\frac{360\pi\eta}{r_i^7\nu^8}\left(1-\alpha^{-7}\right)\right)
		=H(r_i)\exp\left(-\frac{45\eta\lambda^8}{32r_i^7\pi^7}\left(1-\alpha^{-7}\right)\right)\,.
		\label{e-hi}
    \end{equation}
As $r_o\rightarrow\infty$, Eq.(\ref{e-hi}) reduces to 
	\begin{equation}
		H(r_o)=H(r_i)\exp\left(-\frac{45\eta\lambda^8}{32r_i^7\pi^7}\right)\,,
		\label{h2}
	\end{equation}
but the damping effect is reduced by only a little for a shell of finite thickness. For example, $(1-\alpha^{-7})$ is $0.99$ for $\alpha=2$ and is $0.5$ for $\alpha=1.104$.
	To our knowledge, viscous damping of GWs with $r_i\ll\lambda$ has not been studied previously, and Eqs.~(\ref{e-hi}) and (\ref{h2}) are new. In the next sections we investigate some astrophysical implications.
	
	\section{ASTROPHYSICAL APPLICATIONS: GENERAL CONSIDERATIONS}
	\label{consider}
	
	In the next section we will discuss core collapse supernovae (CCSNe), for which viscous damping of GWs can be significant. However, we first make some general comments. The model developed above uses a Minkowski background, rather than Schwarzschild or Kerr for CCSNe. Thus the numerical values that will be obtained should not be regarded as precise statements, but rather as estimates as to when the effect of viscous damping of GWs may be important.
	
	A numerical relativity simulation of the full Einstein and matter field equations with GW extraction far from the source will properly include all effects described above. However, in situations such as CCSNe the complexity of the matter physics necessitates an approximate treatment of GW extraction, normally calculated using a modified quadrupole formula. These models do not consider viscous damping by matter outside the region where the GWs are generated.
	
	Geometric units are used in this paper, but the astrophysical literature reports estimates of the shear viscosity in SI units (kg/m/s = Pa s), or sometimes in cgs units. The conversion requires multiplication by $G/c^3$, where $G$ is the gravitational constant and $c$ is the speed of light; numerically, $G/c^3=2.477\times 10^{-36}$s/kg.
	
	
	\section{CORE COLLAPSE SUPERNOVAE}
	\label{ccsne}
	
	CCSNe have been identified as candidates of sources of detectable GWs. 
	Whilst binary black hole (BBH) and binary neutron star (BNS) mergers are currently the only GW events picked up by LIGO and VIRGO, supernovae are expected to produce, under certain conditions, GWs detectable by the current generation of interferometers or those on the horizon. 
	For now, all detection of supernovae have been confined to electromagnetic detection. The GW signal from a supernova event would be different (but not altogether so) from the characteristic signal of a BBH merger or BNS merger.
	
	Photons originate at the outer edge of a star and hence provide only limited information on the interior regions. The detection of GWs which are the result of the aspherical motion of the inner regions will provide a wealth of information on these regions and the mechanism leading to the supernova explosion, where all the four fundamental forces of nature are involved. 
	
	Whilst the central engines and inner regions of CCSNe have yet to be fully understood, there exist several studies of their progenitors and the subsequent evolution and detection ~\cite{Muller:2020ard,Abdikamalov:2020jzn,Woosley02}.
	For stars of mass larger than 8$M_\odot$, evolution normally proceeds through several stages of core burning and then to core collapse once nuclear fusion halts when there are no further burning processes to balance the gravitational attraction. Typically, these cores are iron cores, with the critical mass signalling the onset of core collapse ranging from $1.3M_{\odot}$ to $1.7M_{\odot}$. The core breaks into two during the collapse, with the inner core 
	of $0.4M_{\odot}$ to $0.6M_{\odot}$ in sonic contact and collapsing homologously and the outer core collapsing supersonically. 
	The inner core reaches supranuclear densities of $\sim 2 \times 10^{14}$gm/cm$^{3}$ where the nuclear matter stiffens, resulting in a bounce of the inner core. The resulting shock wave is launched into the collapsing outer core. However, the shock loses energy to dissociation of iron nuclei, stalling  at ${\sim} 150\,\mathrm{km}$ within ${\sim} 10\,\mathrm{ms}$ after formation. 
	Many computationally demanding simulations exist 
	~\cite{Andresen:2016pdt,Andresen:2018aom,Radice:2018usf} for generation of GWs from CCSNe.
	
	The anticipated GW signal from CCSNe, is normally described by four phases. Initially there is the  convection signal. This is followed by a quiescent phase. The third phase is driven by the standing-accretion-shock instability (SASI) and is also referred to as the neutrino convection phase. Finally, there is the  explosion phase.
	\subsection{General-relativistic simulations of core-collapse supernovae}
	\label{NR}
	
	There has been a steady increase within the numerical relativity community of simulations of the evolution of CCSNe and we summarise some of the results of recent efforts givng some of the important parameter values.
	
	Astrophysics with CCSNe GW signals in the next generation of GW detectors is discussed in ~\cite{Roma:2019kcd} using the Supernova Model Evidence Extractor (SMEE) to capture the main features of GW signals from CCSNe  using numerical relativity (NR) waveforms to create approximate models. These include features in the GW signal that are associated with g-modes and the standing accretion shock instability,  and testing SMEE’s performance using simulated data for planned future detectors, such as the Einstein Telescope, Cosmic Explorer, and LIGO Voyager. In third generation detector configurations, it was found that about 50\% of neutrino-driven simulations were detectable at 100 kpc, and 10\% at 275 kpc. 
	
	Scheidegger et al.~\cite{Scheidegger:2010en} produced 25 gravitational waveforms from 3D magnetohydrodynamic (MHD) core-collapse simulations of a $15\,M_{\odot}$ zero age main sequence star (ZAMS) progenitor star.  They use a variety of rotation values from non-rotating to rapidly-rotating. Rotation leads to a large spike at core-bounce in the plus polarization only. The simulations are short duration as they were stopped up to 130\,ms after the core bounce time.
	
	M\"uller et al.~\cite{Muller:2011yi} carried out 3D neutrino-driven CCSNe simulations of non-rotating stars. The waveforms have emission due to both SASI and g-modes.  The GW signals extend to 1.3\,s after core bounce, with the strongest GW emission in the first 0.7\,s after the core bounce time. The M\"uller et al. waveforms do have g-mode emission, however it is relatively slow to develop and rarely approaches 300\,Hz~\cite{Muller:2011yi}.
	Simulations were performed with the general relativistic neutrino hydrodynamics code Vertex-CoCoNuT~\cite{Mueller2010_umlaut}.
	
	Andresen et al.~\cite{Andresen:2016pdt} also carried out 3D neutrino-driven CCSN simulations of non-rotating stars. 
	They find that the gravitational wave emission is dominated by
	late-time, long-lived convection in the proto-neutron star (PNS). This means 
	that the GW energy produced stems
	mainly from the fluid dynamics within the PNS,
	and not from perturbations of the PNS by fluid dynamics above it.
	Their investigations were confined to rough estimates based on the expected excess power in second- and third generation GW detectors in two bands at low (20$\ldots$250 Hz) and
	high (250$\ldots$1200 Hz) frequency. Third-generation instruments like the Einstein Telescope, however, are expected to detect all of their models at the typical distance of a Galactic supernova
	($\approx$ 10 kpc) and strong GW emitters out to 50 kpc. The GWs were generated within a radius of $10^{4}$\,m to $2.8\times 10^{4}$\,m and if we were to consider this the boundary of the inner radius of our matter shell, then we can take 
	$10^{4}$\,m$\lesssim r_{i} \lesssim 2.8\times 10^{4}$\,m.
	
	Kuroda et al.~\cite{Kuroda:2016bjd} cproduced 3D simulations of a $15\,M_{\odot}$ ZAMS progenitor star using three different equations of state (EOS) and a quadrupole approximation. For the SASI-origin emission, the peak value of GW energy spectrum appears at 129 Hz and reaches
	almost a comparable amplitude to that from g-mode oscillation. 
	It is expected that GWs from Galactic SNe are likely observable even if their progenitors are non-rotating.
	For the SASI-origin emission, $r_{i} \approx 10^{4}$\,m, whilst from the other two models we have
	$10^{4}$\,m$\lesssim r_{i} \lesssim 2\times 10^{4}$\,m.
	
	Yakunin et al.~\cite{Yakunin:2017tus} carried out one general relativistic, multi-physics, 3D simulation of a $15\,M_{\odot}$ ZAMS progenitor star with state of the art weak interactions. 
	They also find that the GW energy produced stems
	largely from the fluid dynamics within the PNS. However, in their model, the dominant emission stems from the convective region itself, rather than from
	the convective overshoot layer above it as in~\cite{Andresen:2016pdt}.
	Their simulation is stopped 450\,ms after core bounce. The strong GW emission starts at $\sim120$\,ms after core bounce when the SASI develops and the emission peaks at a higher frequency of 1000\,Hz due to g-mode oscillations of the PNS surface. Again, $10^{4}$m$\lesssim r_{i} \lesssim 2\times 10^{4}$\,m. 
	
	Powell et al.~\cite{Powell:2018isq} generated two neutrino-driven simulations in 3D down to the innermost 10\,km to include the PNS convection zone in spherical symmetry. The first simulation is the explosion of an ultra-stripped star in a binary system simulated from a star with an initial helium mass of $3.5\,M_{\odot}$. The ultra-stripped simulation ends at 0.7\,s after core bounce. The second is a single star with a ZAMS mass of $18\,M_{\odot}$, which was simulated up to 0.9\,s after core bounce. Both models have peak GW emission between 800\,Hz and 1000\,Hz due to g-mode oscillations of the PNS surface and both models have $r_{i} \approx  10^{4}$\,m.
	
	Shibagaki et al.~\cite{Shibagaki:2020ksk} considered GW and  neutrino signals from  full general relativistic 3-D hydrodynamics simulations of non-rotating and rapidly
	rotating stellar core-collapse. They find that the GW and  neutrino signals would be simultaneously detectable by the current generation detectors up to $\approx$ 10 kpc. Their findings indicate that the joint observation of GWs and neutrinos is indispensable for
	extracting information on the PNS evolution preceding black hole formation. 
	In the non-rotating case, $\Omega_{0}=0$ rad s$^{-1}$, the GWs are generated within a radius of  	 $10^{4}$\,m with a frequency of 200-300\,Hz.
	In the $\Omega_{0}=1$ rad s$^{-1}$ case, $r_{i}\sim 10^{4}$\,m, and the frequncy of GWs are between 300-400\,Hz.
	The $\Omega_{0}=2$ rad s$^{-1}$ model produces GWs within a frequency range of 400-800\,Hz also within a 10km core. 
	
	A summary of the results is given in Table~\ref{parameter}.
	\begin{table}[th]
		\begin{tabular}{lll}
			\hline
			Reference  
			&$r_i$ [m] &Frequency [Hz]\\
			\hline
			\hline
			Roma~\cite{Roma:2019kcd}	
			&& 96 - 1000 \\
			\hline
			Scheidegger~\cite{Scheidegger:2010en}	
			&&317 - 935 \\
			\hline
			M\"{u}ller~\cite{Muller:2011yi}	
			&&130 - 1100 \\
			\hline
			Andresen~\cite{Andresen:2016pdt}
			&$10^{4} - 2.8\times 10^{4}$&100 - 700\\
			\hline
			Kuroda~\cite{Kuroda:2016bjd} &$10^{4} - 2\times 10^{4}$& 100 - 671\\
			\hline
			Yakunin~\cite{Yakunin:2017tus}	&$10^{4} - 2\times 10^{4}$& 200 - 600\\
			\hline
			Powell~\cite{Powell:2018isq}	  &$10^{4}$&800 - 1000\\
			\hline
			Shibagaki~\cite{Shibagaki:2020ksk}	
			&$10^{4}$&200 - 800\\
			\hline
			\hline
		\end{tabular}
		\caption{Parameter values from various references}
		\label{parameter}
	\end{table}
	
	\subsection{The viscosity for a shell surrounding a CCSNe}
	\label{viscosity} 
	
	The viscosity in the core collapse environment has received only a little attention in the literature. Potential mechanisms for viscosity have been explored in~\cite{Thompson2005}, and include  neutrino viscosity, turbulent viscosity caused by the magnetorotational instability (MRI), and turbulent viscosity by entropy and composition-gradient-driven convection. The MRI was found to be the
	most effective dominating the neutrino viscosity by 2 to 3 orders of
	magnitude. Within the PNS,  the authors~\cite{Thompson2005} find that the  MRI will operate and dominate the viscosity even for the slowest rotators considered.
	
	Fig. 5 (left) of~\cite{Thompson2005} plots values of the kinematic viscosity coefficient due the MRI against radius. In the region of the shell, say between 10 and 30\,km, it varies in the range 
	$10^{12}$ to $10^{14}$\,cm$^{2}$s$^{-1}$. Multiplication by the density, about $10^{12}$\,gm/cm${}^3$, gives the dynamic viscosity
	at $10^{23}$ to $10^{25}$\,kg/m/s. Ref.~\cite{Spruit:2001tz} found values consistent with these magnitudes. Values for $\eta$ in neutron star material are discussed in~\cite{Kolomeitsev15} (e.g., see Fig. 25), and can be as high as
	$10^{22}$\,kg/m/s.
	
	

	\subsection{The damping effect on GWs emanating from CCSNe}
	\label{damping}
	\begin{figure}[ht]
		\includegraphics[width=1\columnwidth,angle=0]
		{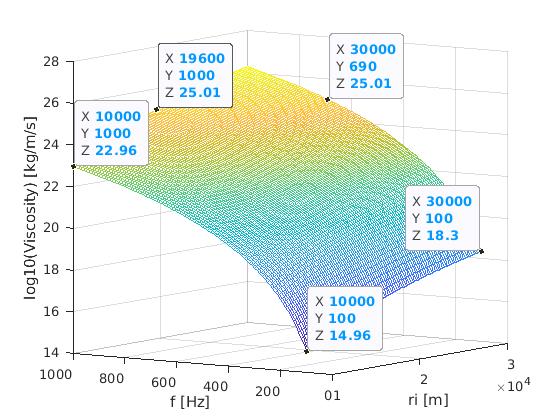}
		\caption{The figure plots the inner radius of the shell ($r_{i}$) on the x-axis (in m), the GW frequency (f) on the y-axis (in Hz), and $\log_{10}(\eta)$ on the z-axis, with $\eta$ in kg/m/s. The surface plotted has damping factor 0.5, i.e. $H(r_{o})/H(r_{i})=0.5$. Some points on the surface are labelled with their coordinate values, indicating which parts of the surface have $\eta>10^{25}$\,kg/m/s..}
		\label{mesh}
	\end{figure}
	
	Putting together the results from the prevous two subsections, we consider the scenario of a CCSNe with GW emission in the frequency range 100 to 1000\,Hz, the inner radius $r_i$ of the matter surrounding the GW source in the range $10^4$ to $3\times 10^4$\,m, and fluid viscosity $\eta$ with a maximum value of $10^{25}$\,kg/m/s. Since $r_i$ is much smaller than the wavelength of the GWs, we can use Eq.~(\ref{h2}).
	Fig.~\ref{mesh} plots the inner radius of the shell ($r_{i}$) on the x-axis (in m), the GW frequency ($f$) on the y-axis (in Hz), and $\log_{10}(\eta)$ on the z-axis, with $\eta$ in kg/m/s. The surface plotted has damping factor 0.5, i.e. $H(r_{o})/H(r_{i})=0.5$. Values of $\eta$ a little above the surface would lead to (almost) complete damping, and those a little below would lead to (almost) no damping.
	
	The figure shows that, except in the case that both $f$ and $r_i$ are towards the top of their ranges, GW viscous damping is expected to be significant.
	%
	\subsection{Detection of GWs from CCSNe}
	\label{detect}
		\begin{figure}[ht]
		\includegraphics[width=1\columnwidth,angle=0]{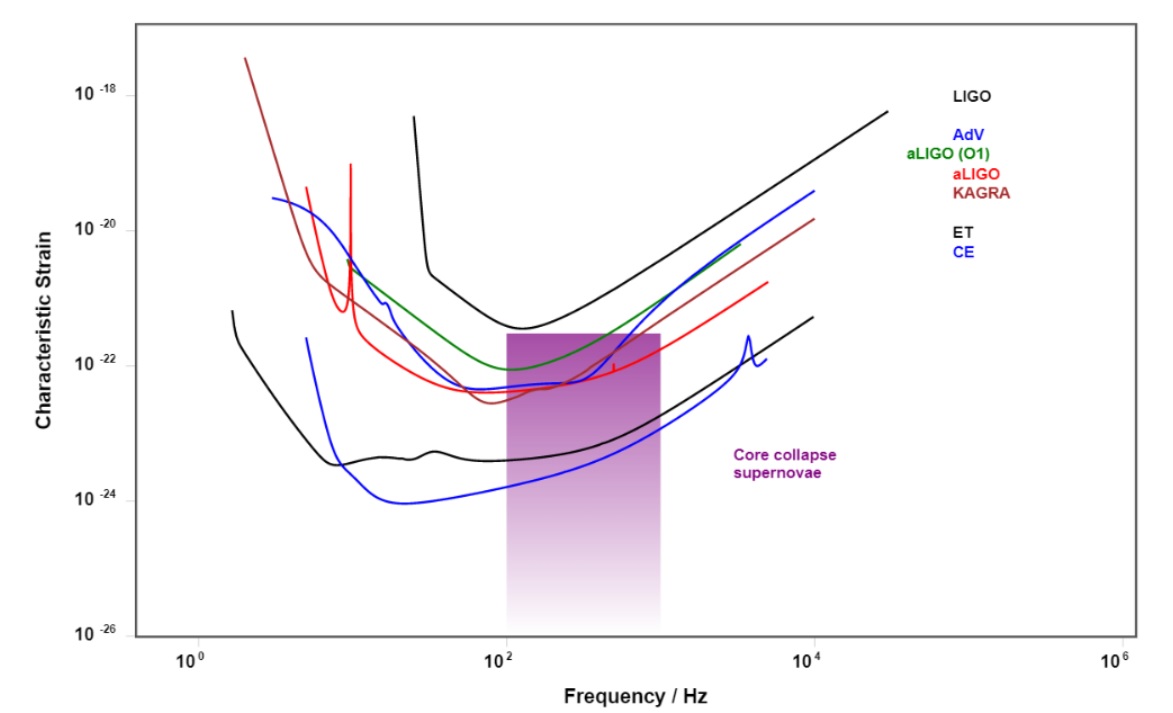}
		\caption{The signal range for CCSNe and sensitivities for various gravitational wave detectors, past, current and future for a source at 300kpc. Figure produced using the web-server reported in~\cite{Moore:2014lga}.}
		\label{spectrum2}
		\end{figure}

	The frequency range of CCSNe and the sensitivity of various detectors is illustrated in Fig.~\ref{spectrum2}. The expected frequency range of GWs from CCSNe lie between $10^{2}$ and $ 10^{3}$Hz, corresponding to a wavelength range of between $3 \times 10^{6}$ and $3 \times 10^{5}$m.  The expected GW magnitude from CCSNe fell outside the sensitivity of LIGO. However, the range fell within the sensitivity of Advanced Ligo, on its first observation run aLIGO (O1). No GWs ascribed to CCSNe were detected on this run or on the subsequent runs, O2 and O3. As the sensitivity improves with each run, towards full design sensitivity, it is hopeful that the increased cachement region leads to the first detection of GWs from CCSNe.
	\section{Primordial GWs}
	\label{s-pGWs}
	
	\subsection{Background}
	
	In our previous work \cite{Bishop:2019eff}, we showed that the thin dust shell approximation can be extended to a thick shell by integrating over multiple shells and from this, we can model the dust-only effect on GWs in cosmology. Due to the extremely low density of the matter dominated Universe, the effect from astrophysical events was, however, found to be negligible. We now consider early Universe epochs where extreme temperatures and density conditions are expected. Primordial gravitational waves (pGWs) are of considerable interest in these epochs since direct observations from electromagnetic radiation are limited to the region after the surface of last scattering, around 370,000 years after the Big Bang, and the expected low interaction of GWs with matter then provides information from very early times. With detection and interpretation of these waves, theories of the early Universe, such as inflation, can then become accessible to direct observation. Similarly, theories of high-energy physics, that go beyond Earth-bound accelerators can also be tested, e.g., probing matter properties in extreme temperatures. Interpreting pGW signals from their interaction with their environment is therefore highly important.
	
	In early works, such as that of Hawking in 1966 \cite{Hawking66}, it was already shown that shear viscosity, $\eta$, will affect pGWs through damping. These results along with the work of Eckart in the 1940s \cite{Eckart1940-od} were consolidated and further developed into what can be considered the seminal work on primordial dissipative processes by Weinberg in 1971 \cite{Weinberg1971}. Since then, there has been significant progress in understanding the details of early epochs and their physical properties but the work of Weinberg is still relevant as baseline formulae for relativistic viscous effects. Accordingly, we will setup representative early Universe scenarios to illustrate the viscous shell effects on pGW in the extreme conditions expected in the epoch after cosmic inflation ends. While recognising that pGWs generated during inflation are expected to be fundamentally stochastic (see \cite{Caprini_2018}), for our purposes here, we will consider a discrete source of pGWs and reserve extending our work to better model the stochastic nature of pGWs for future research.
	

	\subsection{Cosmological scenario}
	
	In order to illustrate the possible effects of viscosity in cosmology, we consider an early Universe scenario based on quantities derived from the standard model and some classical references. We make use of the Friedman-Lema\^{i}tre-Robertson-Walker (FLRW) geometry for a homogeneous and isotropic
	universe with the metric as
	\begin{align}
		ds^2 = -dt^2 + a^2(t)\gamma_{ij}dx^i dx^j,
	\end{align}
	with $\gamma_{ij}$ being the 3-space metric of maximal symmetry and $a(t)$ the scale factor.
	
	Based on the present time as $t=13.7$Gyr $(=4.32 \times 10^{18}s)$ with $T=2.725$K, we normalise the scale factor to be $a_0=1$ at present. Evolving back in time, the density of radiation and matter were approximately equal at $t=56,000$yr ($1.7662\times10^{12}$s) and $T=9,000K$. Prior to that time but after inflation, the Universe can be treated as a radiation dominated Einstein de Sitter model with $a(t) \propto t^{1/2}$. During inflation the expansion is exponential with $a(t) \propto \exp(\mathbf{H}t)$ where $\mathbf{H}$ is the Hubble constant -- we use this notation to avoid confusion with the rescaled GW strain $H$.
	
	At the end of inflation at $t=10^{-32}$s and $T=10^{27}$ to $10^{28}K$. According to \cite{Misner1973-mq} (Section 28.1), $T$ is proportional to $1/a(t)$. Thus, at the end of inflation, we take $a=10^{-27}$. The inverse of the Hubble rate does not change during inflation and is taken as $1/\mathbf{H} (=a/\dot{a})= 10^{-35}s$, which leads to the horizon scale as $c/\mathbf{H} = 3 \times 10^{-27}m$. 

Current ground-based GW detectors operate, approximately, in the frequency range $10$ to $10^3$Hz; pulsar timing arrays and the planned satellite system LISA will extend the lower limit to about $10^{-9}$Hz. Thus searches for pGWs will cover wavlengths in the range $3 \times 10^5$m to $3 \times 10^{17}$m. Since wavelength scales as $a(t)$, this corresponds to a wavelength range at the end of inflation of $3 \times 10^{-22}$ to $3 \times 10^{-10}$m.

	From Weinberg 1971 \cite{Weinberg1971}, rewritten with $c\ne 1$, we get
	\begin{align}
		\tau=\left( \frac{16\pi G\eta}{c^2} \right)^{-1} ,
	\end{align}
	for graviton interaction, where $\tau$ is the particle mean free time and further
	\begin{align}
		\eta=\frac{4}{15}aT^4\tau/c ,
	\end{align}
	again rewritten with $c\ne 1$. Then eliminating $\tau$ from the two equations, we get
	\begin{align}
		\eta^2=\frac{aT^4c}{60\pi G} .
	\end{align}
	When $T=10^{27}$K,
	\begin{align}
		\eta=3.68 \times 10^{58} \, \mathrm{kg/m/s}
		\label{e-eta-W}
	\end{align}
	This value of $\eta$ is very large, and even when multiplied by $G/c^3$ to convert to geometric units, we get a value of $9.09 \times 10^{22}$m${}^{-1}$.
	
	If we consider a thin shell scenario where  $r_i,r_o$ are much larger than the wavelength $\lambda$, using Eq. (\ref{h1}) with $r\nu \gg 1$ and rewriting in terms of $t$, we have
	\begin{align*}
		\frac{dH}{dr}=-8\pi\eta H \rightarrow \frac{dH}{dt}=-8\pi\eta c H .
	\end{align*}
	Now, $\eta$ behaves as $T^2$, i.e. as $1/a^2$ and further, we are in the radiation dominated era and $a$ behaves as $t^{1/2}$. Thus $\eta$ behaves as $1/t$ and we have
	$dH/dt = - 8 \pi \eta_i c H t_i/t$ with $\eta_i=9.09 \times 10^{22}$ and $t_i=10^{-32}$.
	Let $A=8 \pi \eta_i c t_i =6.9$, then integrating
	\begin{align}
		\frac{dH}{H}=-A \frac{dt}{t}
	\end{align}
	we get
	\begin{align}
		H_o=H_i \left(\frac{t_i}{t_o} \right)^{A} .
	\end{align}
For example, suppose that
\begin{equation}
t_o=2t_i\;\;\mbox{then}\;\; H_o=0.009\,H_i\,.
\end{equation}
Although it is believed that after inflation the universe entered a quark-gluon phase and so behaved as an almost ideal fluid, this example shows that if $\eta$ is as large as the value in Eq.~(\ref{e-eta-W}) then the damping effect is so rapid that it may occur before the quark-gluon phase is reached.
	
	Otherwise, considering that the wavelength is in the range of $3 \times 10^{-22}$ to $3 \times 10^{-10}\,$m while the horizon scale is $c/H = 3 \times 10^{-27}$\,m, the case for $r_i,r_o\ll\lambda$ can be justified and Eq. (\ref{h2}) becomes applicable. Using the same formulation, significant damping can then occur for even small values of $\eta$. As opposed to the first scenario where the value of $\eta$ dominates damping, this effect is present in geometries where the $(\lambda/r_i)$ ratio is large, which amplifies the value of the negative exponent in Eq. (\ref{h2}).
	
	As an example, suppose that  $r_i=3 \times 10^{-27}$\,m (a pGW source the size of the horizon scale) with $\lambda=3 \times 10^{-22}$\,m (the lower wavelength limit), then $H_o=0.5 H_i$ for $\eta=2.0\times 10^{25}$\,kg/m/s. This value is much smaller than that of Eq.~(\ref{e-eta-W}), and is of the same order as values considered for CCSNe. The timescale of the damping is $10^{-35}$\,s, which is much shorter than the timescale of inflation.

	\section{SUMMARY AND CONCLUSIONS}
	\label{summary}
	
	We have investigated, using the Bondi-Sachs form of the Einstein equations linearized about Minkoski, the effect of viscosity on GW propagation. A general expression for the damping effect, Eq.~(\ref{e-h0}), was found, which reduces to a known result, Eq.~(\ref{h1}), in the case $\lambda\ll r_i$. However, when $\lambda\gg r_i$, Eq.~(\ref{h2}) applies, which is a novel result. In this case, viscous damping of GWs can be astrophysically important since the effect includes the factor $[\lambda/(\pi r_i)]^7$ which can be large.
	
	The paper then looked at the astrophysical application of CCSNe, but first noting that the model obtained may not be directly applicable so the results should be regarded as estimates rather than as precise predictions.
	
	A number of CCSNe models have been proposed, and it was found that in many cases significant viscous damping of GWs was predicted. GW generation in CCSNe involves a number of different physical processes, each with GW output at a different frequency. It may be that a GW observation of a CCSNe event would see only the higher frequencies, with the lower ones completely damped out; such an observation could be used to constrain a combination of $\eta$ and $r_i$.
	
	We further considered pGWs generated in the early Universe during inflation where the viscous shell model can be applicable. As illustration, a high viscosity configuration was evaluated where $r_i,r_o \gg \lambda$, which resulted in significant damping. A further example considered, was the case $r_i \ll \lambda$. In this scenario, complete damping is predicted for even low viscosity values when the $(\lambda/r_i)$ ratio is high. While these are rather speculative scenarios, we have illustrated that significant pGW damping is possible within reasonable parameter ranges when the combination of geometry and physics allow for viscous shell modelling. On the other hand, if pGWs are detected, the results obtained here would constrain certain physical parameters of the early Universe: the value of the viscosity $\eta$ would need to be significantly lower than that in Eq.~(\ref{e-eta-W}); and there would be a wavelength-dependent constraint on a function of $r_i$ and $\eta$.
	

	%
	\section*{Conflict of interest}
	The authors declare that they have no conflict of interest.
	
	\appendix
	
	\section{Computer algebra scripts}
	The computer algebra scripts used in this paper are written in Maple in plain text
	format, and are available as Supplementary Material. The output file may
	be viewed using a plain text editor wth line-wrapping switched off.
	
	The file driving the calculation is {\texttt ViscousShell.map}, and it takes input from the files {\texttt gamma.out, initialize.map, lin.map} and {\texttt ProcRules.map}; the output is in {\texttt ViscousShell.out}. The Maple script is adapted from that reported in our previous work~\cite{Bishop:2019eff}. It first calculates and outputs the metric coefficients given in Eqs.~(\ref{e-pert}), and it then substitutes this solution into the Einstein equations and checks that they are all satisfied. The next step is to use the matter conservation conditions for dust, obtaining Eqs.~(\ref{e-V}) for the fluid velocity field; the script also checks that the solution obtained satsifies the geodesic conditions. Next, the shear tensor $\sigma_{ab}$ and the expansion $\theta$ are found, giving Eqs.~(\ref{e-sig}). Next, we evaluate the expression $-2\eta \sigma_{ab}\sigma^{ab}$, intergrate the result over a spherical shell of radius r and thickness $\delta r$, and find the time average; the result is Eq.~(\ref{e-dE_eta}).

	\bibliographystyle{spphys}
	\bibliography{t_1,aeireferences}
\end{document}